\documentclass[aps,prl,twocolumn]{revtex4}
\usepackage{amsfonts}
\usepackage{mathrsfs}
\usepackage{amsmath}

\begin{document}
\title{Reply to the Comment of X. Ji on
``Do gluons carry half of the nucleon momentum?''
[PRL 103:062001 (2009)]}
\author{Xiang-Song Chen,$^{1,2,3}$
Wei-Min Sun,$^3$ Xiao-Fu L\"{u},$^2$ Fan Wang,$^3$  and T.
Goldman$^4$} \affiliation{$^1$Department of Physics, Huazhong
University of Science and Technology, Wuhan 430074, China\\
$^2$Department of Physics,
Sichuan University, Chengdu 610064, China\\
$^3$Department of Physics, Nanjing University, CPNPC, Nanjing
210093, China\\
$^4$Theoretical Division, Los Alamos National Laboratory, Los
Alamos, NM 87545, USA}
\date{\today}

\begin{abstract}
We affirm that the proper momentum defined in [PRL 103:062001
(2009)] does respect exact gauge symmetry and is as measurable as
the kinetic momentum. The physical part of the gauge field is also
as measurable as the electromagnetic field. The Comment of Ji
[arXiv:0910.5022] is due to a misunderstanding of our work, and a
typical confusion of our gauge-invariant formalism with the specific
Coulomb-gauge calculation.
%\pacs{11.15.-q, 14.20.Dh, 12.38.-t, 12.20.-m}
%11.15.-q Gauge field theories
%14.20.Dh Protons and neutrons
%12.38.-t Quantum chromodynamics
%12.20.-m Quantum electrodynamics
\end{abstract}
\maketitle

The key obstacle to gauge-invariant construction of some physical
quantities (e.g., the gluon spin) is the inevitable involvement of
the gauge-dependent field $A^\mu$. Recently, we proposed a method to
decompose the gauge field: $A^\mu \equiv A^\mu_{\rm phys}+A^\mu_{\rm
pure}$ \cite{Chen09,Chen08}. The aim is that $A^\mu_{\rm phys}$ will
be a physical term which is gauge-covariant and always vanishes in
the vacuum, and $A^\mu_{\rm pure}$ is a pure-gauge term which solely
carries the gauge freedom and contributes nothing to the
electromagnetic fields, $\vec E$ and $\vec B$. Equipped with the
separate $A^\mu_{\rm phys}$ and $A^\mu_{\rm pure}$, a naively
gauge-dependent quantity (such as the gluon spin $\vec S =\vec
E\times \vec A$) can easily be rescued to be gauge-invariant, simply
by replacing $A^\mu$ with $A^\mu_{\rm phys}$, and by replacing the
ordinary derivative with the pure-gauge covariant derivative
constructed with $A^\mu_{\rm pure}$ instead of $A^\mu$.

In his Comment \cite{Ji09}, Ji claimed that the physical field
$A^\mu_{\rm phys}$ ``is never an observable in electromagnetism as
$\vec E$ and $\vec B$ are". Here we show that this claim is due to a
misunderstanding of our method. Particularly, Ji did not understand
that in our formalism $A^\mu_{\rm phys}$ can be expressed entirely
in terms of $\vec E$ and $\vec B$, therefore is as measurable as
$\vec E$ and $\vec B$ are.

Mathematically, the well-defined separation $A^\mu \equiv A^\mu_{\rm
phys}+A^\mu_{\rm pure}$ which we propose is an unambiguous
prescription for constructing $A^\mu_{\rm phys}$ and $A^\mu_{\rm
pure}$ out of a given $A^\mu$. The properties (especially, gauge and
Lorentz transformations) of $A^\mu_{\rm phys}$ and $A^\mu_{\rm
pure}$ are inherently determined via their mathematical expressions
in terms of $A^\mu$. In the Abelian case which Ji concentrated and
commented on, the defining equations for the separation $A^\mu
\equiv A^\mu_{\rm phys}+A^\mu_{\rm pure}$ are
\begin{subequations}
\label{A}
\begin{eqnarray}
&&F^{\mu\nu} _{\rm pure} \equiv \partial ^\mu A^\nu_{\rm
pure}-\partial^\nu
A^\mu_{\rm pure}=0, \label{A1}\\
&&\vec \nabla \cdot \vec A_{\rm phys} =0, \label{A2}
\end{eqnarray}
\end{subequations}
with the boundary conditions that, for a finite physical system,
$A^\mu_{\rm phys}$ approaches zero at spatial infinity (as does the
field strength $F^{\mu\nu}$), and $A^\mu_{\rm pure}=A^\mu -
A^\mu_{\rm phys}$ approaches $A^\mu$. These equations and boundary
conditions lead to familiar solutions. But Ji did not understand
that the more convenient procedure is to first solve $A^\mu_{\rm
phys}$, and then obtain $A^\mu_{\rm pure}$ as $A^\mu -A^\mu_{\rm
phys}$. To this end, Eqs. (\ref{A}) can be arranged into the more
transparent form:
\begin{subequations}
\label{A'}
\begin{eqnarray}
&&\vec \nabla \times \vec A_{\rm phys}=\vec  B ,\\
&&\vec \nabla \cdot \vec A_{\rm phys} =0 , \\
&&\vec \nabla A^0_{\rm phys}=-\partial _t \vec A_{\rm phys} -\vec E
,
\end{eqnarray}
\end{subequations}
which indicate clearly that $A^\mu _{\rm phys}$ is solely determined
by $\vec E$ and $\vec B$. The explicit solution is
\begin{subequations}
\label{As}
\begin{eqnarray}
\vec A_{\rm phys}=-\vec \nabla \times \frac{1}{\vec \nabla ^2} \vec
B =\int d^3 x' \frac {\vec B(\vec x',t)\times (\vec x-\vec x')}
{4\pi |\vec x-\vec x'|^3} \\
A^0_{\rm phys}=\int _\infty^x dx'^i (\partial_t (\vec \nabla '
\times \frac{1}{\vec \nabla '^2}\vec B)^i  -E^i) \label{As2}
\end{eqnarray}
\end{subequations}
(Here the index $i$ takes any value of 1 to 3, and is not summed.)
These explicit expressions clearly justify that $A^\mu_{\rm phys}$
is gauge invariant and physical in the usual sense. $A^\mu_{\rm
phys}$ vanishes as $F^{\mu\nu}=0$. Since $A^\mu_{\rm phys}$ is
entirely expressed in terms of $\vec E$ and $\vec B$, according to
the usual understanding $A^\mu_{\rm phys}$ is as measurable as $\vec
E$ and $\vec B$ are. In consequence, the proper momentum $\vec P_q
\equiv -i \vec \nabla -q\vec A_{\rm pure}$ is as measurable as the
well-known kinematic momentum $\vec \pi \equiv -i\vec \nabla -q\vec
A=\vec P_q -q\vec A_{\rm phys}$, which relates to a charged
particle's velocity in an electromagnetic field.

Ji made a very weird critique that our formalism is ``non-local'' in
the sense that an instantaneous integration is involved in
Eqs.~(\ref{As}). We do not see what's wrong with such kind of
``non-locality'' which in fact is frequently needed: The total
charge $Q(t)$ in a region at an instant $t$ is defined
``non-locally'' in terms of the instantaneous local charge density:
$Q(t)=\int d^3x \rho (\vec x,t)$. The world population $N(t)$ must
also be counted ``non-locally'': $N(t)=\int d^3x n(\vec x,t)$, with
$n(\vec x,t)$ the instantaneous local population density.
Accordingly to Ji's standard, even the electric charge and the world
population are unmeasurable and unmeaningful!

A more resembling ``non-local'' example is the macroscopic
electromagnetic fields $\langle {\vec E} \rangle$ and $\langle {\vec
B} \rangle$ in a medium, which are defined by ``non-local'' average
of the microscopic field $\vec E$ and $\vec B$ \cite{Jack99}. E.g.,
$\langle {\vec E} (\vec x, t) \rangle =\int d^3 x' f(\vec x-\vec x
') \vec E(\vec x ',t)$, with $f(\vec x)$ a suitable sampling
function. $A^\mu_{\rm phys}(\vec x,t)$ and $\langle {\vec E} (\vec
x, t) \rangle$ are both defined at a local point $(\vec x,t)$, but
they are expressed ``non-locally'' in terms of $\vec E(\vec x,t)$
and $\vec B(\vec x,t)$. Such ``non-local'' construction is
physically meaningful and useful, and nothing to worry about. One
should be alert to non-locality only when it might lead to violation
of micro-causality \cite{note}. This does not happen here to
$A^\mu_{\rm phys}(\vec x,t)$ and $\langle {\vec E} (\vec x, t)
\rangle$: When quantized, the commutators involving them vanish
rapidly at space-like intervals significantly larger than the
characteristic wavelength.

Our field separation does not interfere with any specific gauge
choice. It can be regarded as just a mathematical prescription to
construct $A^\mu_{\rm phys}$ and $A^\mu_{\rm pure}$ in terms of the
full $A^\mu$, to which any gauge condition can be assigned. Ji made
a typical confusion of Eq.~({\ref{A2}) with the Coulomb gauge
condition $\vec \nabla \cdot \vec A =0$, especially because in
practice the Coulomb gauge does simplify the expressions of
$A^\mu_{\rm phys}$ and $A^\mu_{\rm pure}$ in terms of $A^\mu$, thus
make calculations easier.

From Ji's critique that our field separation is not Lorentz
symmetric and does not lead to proper Lorentz transformation, we can
see that Ji actually does not understand correctly what Lorentz
symmetry and proper Lorentz transformation are. Lorentz symmetry
just requires that the basic physical laws, as expressed by the
basic field equations, take the same form in different Lorentz
frames. And a proper Lorentz transformations is the one that leaves
invariant the basic field equations. It is nice but not required
that all quantities transform in the standard manners of Lorentz
scalar, vector, etc.. Indeed, massless particles with spin $\geq 1$
necessarily behave in non-standard manners \cite{Wein95}. Concerning
our separation, the defining equations (\ref{A}) do take the same
form in all Lorentz frames. In quantum language, Eqs.~(\ref{A})
commute with all Lorentz generators \cite{Mano87}. $A^\mu_{\rm
phys}$ must not transform as a four-vector, but this non-standard
behavior, as expected for the massless spin-1 particle, is just the
proper Lorentz transformation of $A^\mu_{\rm phys}$ so as to
preserve Eq.~(\ref{A2}) in all Lorentz frames.

To summarize: We are not doing what Ji calls
``reversely-engineered'' gauge symmetry. $A^\mu_{\rm phys}$ and
$\vec P_q=\vec \pi +q\vec A_{\rm phys}$ have exact gauge symmetry in
the usual sense. Very much physics and convenience can be gained by
separating the kinetic momentum into gauge-invariant and measurable
combination $\vec \pi =\vec P_q -q\vec A_{\rm phys}$. It reveals how
much of the kinetic momentum is actually due to interaction with the
gauge field. Moreover, it is $\vec P_q$ rather than $\vec \pi$ that
commutes in two directions, and generates spatial translation for
the Dirac field. The proper quark and gluon momenta in \cite{Chen09}
have exact gauge symmetry and the calculation is complete.
Historically, it took tremendous effort to establish factorization
formulae to measure the quark kinetic momentum. We do not expect
that measuring the proper quark momentum would be easier. On the
other hand, no one has proved, and it is very imprudent and
irresponsible for Ji to claim, that this measurement is impossible.

In closing, we would like to point out several minor but unfortunate
mistakes in Ji's Comment: Typo-like expression ``$A^\mu_{\rm phys}
=A^\mu - \vec A_{\rm pure}$'', and incoherent sentence ``Doesn't
$\vec A_\bot$ describe the physical degrees of freedom and other
components of $A^\mu$ are the pure gauge part?''. Apparently, Ji did
not understand that in our formalism both the physical and
pure-gauge fields have four components, thus he had in mind such
misconception as ``the rest of $A^\mu$ other than $\vec A_\bot$''.
Finally, Ji even miscopied the title of our paper to be ``Does
Gluons Carry Half of the Nucleon Momentum?''.

\end{document}